%Paper: nucl-th/9309025
%From: PICHON@MESIOB.OBSPM.CIRCE.FR
%Date: Fri, 24 Sep 1993 15:15:32 +0002
%Date (revised): Fri, 24 Sep 1993 17:32:03 +0002

\magnification=1200
\tolerance=1000
\pretolerance=500
\hfuzz=5pt
\overfullrule=0pt
\brokenpenalty=5000
\vsize=23.0truecm
\hsize=16.0truecm
\baselineskip=12pt
\newcount\pageun
\pageun=1
\headline={\ifnum\pageno=\pageun \hfil \else
 \ifodd\pageno\rightheadline \else \leftheadline \fi \fi}
\footline={\hfill} % i.e. \nopagenumbers
\def\titrepic{\null}
\def\rightheadline{\tenrm\hfil{\it\titrepic}\qquad\rlap{\bf\folio}}
\def\leftheadline{\tenrm\llap{\bf\folio}\qquad{\it\titrepic\hfil}}
\mathchardef\zzzalpha="010B\def\alpha{\ifmmode\zzzalpha\else$\zzzalpha$ \fi}
\mathchardef\zzzgamma="010D\def\gamma{\ifmmode\zzzgamma\else$\zzzgamma$ \fi}
\mathchardef\zzzdelta="010E\def\delta{\ifmmode\zzzdelta\else$\zzzdelta$ \fi}
\mathchardef\zzztheta="0112\def\theta{\ifmmode\zzztheta\else$\zzztheta$ \fi}
\mathchardef\zzzsigma="011B\def\sigma{\ifmmode\zzzsigma\else$\zzzsigma$ \fi}
\def\cf{{\it cf.\/}~}
\def\ie{{\it i.e.\/}~}
\def\eg{e.g.\ }
\def\viz{{\it viz.\/}~}
\def\s{{\tt s}\ }
\def\p{{\tt p}\ }
\def\r{{\tt r}\ }
\def\keV{\hbox{\rm keV} }
\def\MeV{\hbox{\rm MeV} }
\def\up#1{\raise 1ex\hbox{\sevenrm#1}}
\def\nuc.#1.#2.{\ifmmode \sp{#2}{\rm#1} \else $\sp{#2}{\rm#1}$ \fi}
\def\tnumt#1{\ifmmode\message{Pb-dollards}\else $ - \, #1 \, - $ \fi}

\def\Ep{\ifmmode{E_p}\else$E_p$\ \fi}

\def\subsh.#1.#2.#3.{#1\hbox{#2}\sb{#3/2}}
\def\ref#1{\noindent\null{\up{#1)}}}
%%%%%%%%%%%%%%%%%%%%%%%%%%%%
\null
\vskip 20truemm
\centerline{\bf COMBINATORIAL STUDY OF NUCLEAR LEVEL DENSITY }
\centerline{\bf FOR ASTROPHYSICAL APPLICATIONS }
\bigskip
\bigskip
\centerline{\bf Bernard PICHON}
\medskip
\centerline{ D\'epartement d'Astrophysique Relativiste et de Cosmologie }
\centerline{ DARC - UPR176 du CNRS }
\centerline{ Observatoire de Paris, Section de Meudon }
\centerline{ 92195 Meudon Cedex -- France }
\vskip 4truecm
\noindent{\bf Abstract :  } We present some results and remarks based
on a combinatorial approach of the evaluation of the nuclear level
density.  First, we show that it is possible to extract some reliable
information from the output of the program whose rough data present a
strong statistical fluctuation from bin to bin.  This includes smoothing
and evaluation of the desired quantities.  After some comments about the
spin and parity distributions, we consider the question of the
non-equipartition of parities, mainly, at low energies.  Finally, we
present a simple model to include and test this effect in the
computation of thermonuclear reaction rates.  \par
\vfill
\noindent {\it Accepted for publication in Nucl. Phys. A} \par
\vfill \eject
\null
\bigskip
\beginsection{   1 : INTRODUCTION }\par
\medskip
\indent Nuclear level densities are one of the suitable quantities for
nuclear physics and, especially, for nuclear reaction models (\eg the
statistical Hauser-Feshbach theory). Therefore, they have been
investigated a long time ago, the first attempt, using the infinite
non-interacting Fermi-gas model, being that of Bethe \break in 1936
\ref{1}.  \par
\indent The need to evaluate reliable reactions rates, and thus nuclear
level density, for numerous nuclei arises from the specificity of
nuclear astrophysics which deals with nuclei in the whole chart of
nuclides, towards the proton drip line (for \alpha-p or r-p process),
near both sides of the valley of stability (for \p and \s processes) or
towards the neutron drip line (for the \r- process). In all these cases,
very few reaction rates are measurable experimentaly, especially at the
energies of astrophysical interest. For all the others, only theoretical
predictions are possible. \par
\indent For this purpose, we must develop some {\it ab initio}, but
rather simple, methods (in order to minimize the computational work)
with the minimum number of parameters. This is particularly the case of
the nuclear level density. \par
\bigskip
\indent The aim of this paper is to present a combinatorial method of
computing the level density by enumeration of all the possible
configurations, and to obtain easily a simple law describing the energy
dependence of the level density. Beyond the spin distribution, we focus
our attention onto the parity distribution which, partially, originates
this work. The final intent is to evaluate the various uncertainties,
constraints or difficulties related with the level density problem, and
their impact on reaction rates predictions. \par
\indent Some of the previous attempts are briefly reviewed and commented
in \S 1. Our attempt is depicted in \S 2 with the description of
the model and the preliminary treatments. Some results, including the
determination of the level density parameter and a comparison with the
usual spin distribution function, are shown in \S 3. The problem of
equipartition of the parity distribution for the lowest energy levels is
presented in \S 4. In our conclusions (\S 5), we consider some possible
developments or applications, a few of them (mostly astrophysical ones)
being now in progress. The technical part of our configuration
generator algorithm will be described elsewhere. \par
\bigskip
\indent There are basically two main classes of methods for evaluating
the nuclear level densities. The first methods, and the most used, are
based on the statistical approach via the partition function (see \eg
refs. \ref{2}, \ref{3} for review, or \ref{4}, \ref{5} for
computational programs). The other method is the combinatorial approach
which has been much less investigated, mainly due to the extent of the
computations it requires. Now, let us focus on this last one. \par
\indent To our knowledge the first attempt is due to Hillman and Grover
\ref{6}. They used different sets of single particle levels, the
pairing energies being evaluated within the BCS approximation. In the
same spirit, a recursive method has been deviced by Williams \ref{7},
that represents an interesting alternative way of computing level \break
densities (see \S 2-c). \par
\medskip
\indent Another attempt was done by Ford \ref{8}. He used Woods-Saxon or
Nilsson single particle levels with a new combinatorial algorithm to
generate the configurations. This algorithm works according the
$M-$scheme. This implies special and supplementary assumptions in order
to obtain the $J-$scheme configuration, and thus, the spin distribution
and, more importantly, the energy of the configuration under \break
consideration (see \S 2). \par
\null
\bigskip
\beginsection{ 2 : THE PRESENT ATTEMPT }\par
\medskip
\noindent{\bf a --- Description of the model}\par
\smallskip
\indent As previously said, the combinatorial method is based on a set
of single particle levels, on an algorithm to generate configurations,
and on an evaluation of various physical quantities of each
configuration under study such as energy, parity and spin distribution
(due to degeneracy). \par
\medskip
\indent The results presented here have been obtained with the use of
spherical single particle levels derived within the framework of an
energy density approximation described in ref. \ref{9}. We do not
investigate here the influence on the level density of a change of those
single particle spectra. This question will be examined elsewhere
along with the consideration of deformed single particle spectra. \par
\indent In order to generate all the configurations up to a given maximum
energy (chosen less or equal to $ 30 \ \MeV $), we have adapted Ford's
algorithm to our specific purposes. In fact, the algorithm designed by
Ford explores all the possible configurations by increasing energies.
Hence, it avoids the evaluation of unnecessary physical quantities for
configurations with energy far in excess of the adopted upper limit. \par
\indent The $M-$scheme used in Ford's algorithm \ref{8} to classify the
configurations has been converted into the desired $J-$scheme. This
transformation is complicated by the fact that many
``$M-$configurations'' lead to the same ``$J-$configuration'' (see the
example given in table 1). We have also added to the original algorithm
in order to keep only the first occurrence of a new
``$J-$configuration'' ;  the following ``$M-$configurations'' that lead
to a previous studied ``$J-$configuration''being skipped. The technical
details of our modified algorithm will be presented elsewhere. \par
\smallskip
\indent The pairing is treated following the idea originated by Ford \ref{8}.
Namely, each configuration is inspected subshell by subshell.
In each one, we examine all the possibilities to form pairs with the available
nucleons. Each time that a nucleon pair can be formed, we subtract the
so-called ``pairing energy'' $ E_p $ (as defined by Ford). Such a
procedure looks like a seniority scheme applied subshell by subshell to
the whole configuration. \par
\indent Unfortunately, when considering a non-vanishing pairing energy
\Ep , his algorithm fails. More precisely, it produces an erroneous
spin distribution as can be seen easily from a simple example. Let us
consider the $ (\subsh.1.d.5.) \sp{3} $ configuration with a pure \delta
pairing force $ \Ep \times \delta \sb{J=0} $. With the help of the
Mayer-Jensen'table (see \eg ref. \ref{6}), we find a ``ground'' state $ J =
{5\over2} $ and two ``degenerate'' states $ J = {3\over2} $ and $ J =
{9\over2} $. If we follow Ford's procedure, we obtain all the $ \left(
m_1 m_2 m_3 | M \right) $ states (see table 1). In order to find the
ground state, we collect, as suggested, the $M-$states by pairs (one
$m_i$ and its opposite). Thus, we find the following states and their
corresponding spin (see table 1, last column). Considering that the
number of states with spin $J$ is equal to the number of states with $
M=J $ minus the number of states with $ M=J+1 $ , we obtain a twofold
degenerate ``ground'' state with $ J = {5\over2} $ and an ``excited''
state with $ J = {9\over2} $ (for which the only projections $ M = \pm
{9\over2} $ and $ M = \pm {7\over2} $ exist \dots, the other values of
the magnetic quantum number being counted as spurious in one of the $ J
= {5\over2} $ ``ground'' state). \par
\smallskip
\indent We have considered a number of parametrizations for the
``pairing'' energy \Ep (expressed in $\MeV$) :  \par
\def\boule{\item{$\bullet$}}
\boule $ \Ep = 0 $ \qquad \qquad \qquad \ie no pairing \hfill $(1a)$ \par
\boule $ \Ep = 9.6 A\sp{-1/2} $ \qquad as used by Ford \hfill $(1b)$ \par
\item{} \hfill ref. \ref{8} \par
\boule $ \Ep = \cases{ 7.36 A\sp{-1/3} ( 1 - 8.15 \eta^2 ) & for neutrons \cr
                       7.55 A\sp{-1/3} ( 1 - 6.07 \eta^2 ) & for protons \cr }
$ \hfill $(1c)$ \par
\item{} \hfill ref. \ref{10} \par
\smallskip
\noindent where \ $ \eta = { N - Z \over A } $ \ is the neutron --
proton asymmetry (see also refs. \ref{11}, \ref{12}, \ref{13} and the well
known droplet pairing energy). \par
\smallskip
\indent In all cases, \Ep is identified with the pairing term of a
nuclear mass formula, generally assumed to be equal, with the
appropriate signs, to the back-shift \delta \break (see eq.$(4)$). In
addition, we assume that \Ep does not vary with the excitation energy.
The simplest way to avoid this approximation is to compute an effective
pairing energy \Ep for {\it every} single particle subshell. To do
this, we may consider a (fictitious) nucleus with all subshells filled
up to the subshell under consideration. The pairing energy of {\it
this} subshell can be evaluated from the energy difference between the
configuration energies computed with and without pairing interaction.
Such an improvement is able to take into account the fact that the
pairing energy vanishes at high energy. \par
\indent For a given ``$J-$configuration'', a Mayer-Jensen table (see ref.
\ref{6}) provides the spin distribution for each group of states, sorted
by energy as done by the treatment of the pairing energy. Of course, in
the $ \Ep = 0 $ case, there is only one set of states at the same
energy, and hence the degeneracy of this level is maximum even for the
``ground'' state. \par
\indent Once the proton and neutron levels are obtained, they have to be
convoluted in order to obtain the final distribution of nuclear levels.
For the coherence of our assumptions, no residual $p-n$ interaction has
been considered. \par
\medskip
\noindent{\bf b --- Preliminary treatments}\par
\smallskip
\indent For each energy bin, from $ 0 \ \MeV $ up to $ 30 \ \MeV $
by $ 10 \ \keV $ steps, the algorithm described in the previous
section provides the number of states of given spin (from $ J = 0 \
\hbox{or}\ J = {1\over2} $ to $ J = 60 \ \hbox{or}\ J = {121\over2} $)
with the indication of parity. As we see on figure 1a, there is, as
expected, a strong statistical fluctuation from bin to bin. It is
necessary to treat these data in order to extract a reliable
information, like the level density parameter $a$ (see \S 3-c). \par
\medskip
\indent Let us notice that the aim is to smooth data in order to be able
to extract parameters for some fits (see \S 3-a), and then, using them
to compute the nuclear level density at any energy. The goal is {\bf
not} to smooth data in order to evaluate, by some interpolation, the
density at an arbitrary energy from the smoothed data. In fact, the
observed ripples prevent such a treatment :  for example, the resulting
density might not be a monotonically increasing function of the energy !
Thus, several smoothing methods have been tested \ref{14}. \par
\noindent \tnumt{1} ``The minimum - maximum - median method'' (hereafter
3m-method). Each bin is associated with the set of its neighbors
(typically a few tens on both sides). Each set is sorted in increasing
order and, some extreme points being possibly discarded (for example to
take account of empty bins). In the remaining points, either the first
point (the minimum), or the last point (the maximum) or the middle point
(the median) is selected. This point represents the value of the
smoothed data for the bin under consideration. \par
\indent One of the advantages of this method is its high controlability
on the data treatment. Unfortunately, it requires a lot of
computational time, due to the necessary sorting. Thus, we have turned
our attention to faster methods. For instance : \par
\noindent \tnumt{2} The convolution of the input distribution with the
Fejer's function (\viz $ \propto \left( 1 - (x/n)\sp{2} \right)\sp{2} $)
which guarantees the smoothing to be compatible with the evaluation of
the first derivative. This can be useful when dealing with the
cumulative number of levels, and not directly with the level density. \par
\noindent \tnumt{3} The running mean method based on the convolution of
the input distribution with a rectangular step function. This simple
method has been adopted here (the average being taken over $ 100 \ \keV $).
The result of our smoothing is presented on figure 1b for the
same nucleus as in figure 1a. Notice that the cumulative number of
levels, involving a quadrature, provides a smoother curve than the level
density itself, but the extraction of parameters would be less reliable
and easy. \par
\medskip
\noindent{\bf c --- Other possible models}
\smallskip
\indent As noted previously by various authors (see, for instance,
Hillman and Grover \ref{6} or Ford \ref{8}) the computational time
represents a quite severe restriction to the use of the combinatorial
method. The recursive method proposed by Williams \ref{7} is, in this
respect, very efficient but is not directly applicable to handle the
parity and the spin distributions. \par
\indent Recently, a Monte-Carlo approach has been developped \ref{15}.
Using such a method, one may obtain similar results with some
advantages, for instance, a short computational time which could allow
the use of more sophisticated treatments of pairing. In addition, it
allows also to take into consideration residual interactions between
neutrons and protons. However, there are some disadvantages, such as
the need to test the sampling function, to calibrate results with other
methods, like the full combinatorial method described here. We think
that the Monte-Carlo method will be very suitable for the heaviest
nuclei, where the combinatorial method is still computationally
unworkable, whereas our full combinatorial method represents the best
compromise for moderate heavy nuclei, when the desired number of levels
does not exceed a few millions and when the Monte-Carlo approach may
suffer from lack of statistics. \par
\null
\bigskip
\beginsection{ 3 : RESULTS AND REMARKS ABOUT LEVEL DENSITY}\par
\smallskip
\noindent We have used our model for four sets of nuclei : \hfill\break
\noindent --- \nuc.Ca.40., \nuc.Ti.44., \nuc.Ti.46., \nuc.Cr.48.
and \nuc.Cr.50. , called the ``magical'' series since their study allows
the evaluation of various physical parameters and their evolution with
the distance from a doubly magic nucleus, \hfill\break
\noindent --- an odd $A$ isobaric series : \nuc.K.47., \nuc.Ca.47.,
\nuc.Sc.47., \nuc.Ti.47. and \nuc.V.47. , \hfill\break
\noindent --- an even $A$ isobaric series : \nuc.K.50., \nuc.Ca.50.,
\nuc.Sc.50., \nuc.Ti.50., \nuc.V.50., \nuc.Cr.50., \nuc.Mn.50. and
\nuc.Fe.50. , \hfill\break
\noindent --- for comparison purposes, the nuclei already studied by
Ford, \ie \nuc.Fe.56., \nuc.Co.59., \break \nuc.Ni.60., \nuc.Ni.62.,
\nuc.Cu.61. and \nuc.Cu.63. . \par
\noindent The two isobaric series have been adopted for the purpose of
comparison with the results of Pieper \ref{16}. \par
\indent For all these nuclei, the distributions of spins and parities
have been evaluated up to $ 30 \ \MeV $. As expected, the
computational time varies much from a nucleus to another~:  the computer
time needed varies typically (for a VAX8600) from 40 mn up to 1 day.
This indeed represents a very drastic limitation of the use of such a
method. \par
\medskip
\noindent{\bf a --- Comparison with nuclear level density formulae}\par
\smallskip
\indent After smoothing the results derived from the algorithm described
in \S 2-b, we can try to extract some parameters that appear in
classical level density formulae. Here, we consider the so-called
``Bethe back-shifted Fermi gas formula'' (see, for instance, ref. \ref{8}
appendix B) :
$$ \rho(E) = { 1 \over \sqrt{2\pi} \sigma } { \sqrt{\pi} \over 12
a\sp{1/4} } { \exp ( 2 \sqrt{aU}) \over U \sp{5/4} } \eqno(2) $$
\noindent where $ \rho(U) $ is the level density at energy
$ U = E - \delta $ , \delta being the back-shift, introduced
to take into account the differences between nuclei according
to evenness or oddness of $N$ and $Z$, $a$ is the so-called
``level density parameter'', and \sigma is related to the
spin-distribution. \par
\indent If the classical energy dependence of \sigma
(see \S 3-e, eq.$(11)$) is adopted, eq.$(2)$ transforms into :
$$ \rho(E) \propto U \sp{-3/2} \ \exp \left( \sqrt{2aU} \right)
\eqno(3) $$
\indent We have investigated the change of the exponent $3/2$ of $U$ in
eq.$(3)$ by $ 3/2 + x $, where $x$ is constant. It appears that
only $ |x| < 0.1 $ is compatible with data, so that the conventional
eq.$(3)$ has been retained. \par
\medskip
\noindent{\bf b --- Comparison between some determinations of the
back-shift \delta }\par
\smallskip
\indent First, the determination of \delta proves to be very difficult
from our data and with our treatment. In fact, the cumulative number of
states $ N(E) $ (that minimizes the statistical
fluctuations, see figure 1b) cannot provide any information on this
parameter, since, whatever \delta , $ N(E=0) = 0 $. One has thus to
work with the level density itself. Then, the main drawback is that the
large statistical fluctuations prevents a reliable determination of
\delta. In such conditions \delta is assumed to be equal to the pairing
energy of the liquid drop nuclear mass formula, so that :
$$ \delta = { 1 \over 2 } \ \left[ (-1)\sp{N} \Ep(n) + (-1)\sp{Z} \Ep(p)
\right] \eqno(4) $$
\noindent where $ \Ep(n) $ and $ \Ep(p)$ are the neutron and proton
pairing energies. This assumption is coherent with our treatment of the
pairing. The results we present in this paper are determined with the
pairing energies given in eq.$(1c)$. \par
\indent For comparison, we consider the phenomenological determination
\ref{17} : \par
$$  \delta = \Delta - 5.6 A\sp{-0.28} \eqno(5) $$
\noindent where \par
$$ \Delta = { 1 + (-1)^N \over 2 } \Delta_n +
{ 1 + (-1)^Z \over 2 } \Delta_p \eqno(6) $$
\noindent $\Delta_n$ and $\Delta_p$ being the experimental neutron
and proton pairing energies (see ref. \ref{18}). \par
\indent Table 2 compares the values of \delta extracted from eq.$(5)$
with those determined in this work (see eq.$(4)$). The observed
discrepancies do not exceed $ 0.5 \ \MeV $ from the average value, and
can be considered as quite negligible with respect to the uncertainties
arising from the determination of the level density parameter $a$
(see \S 3-d). \par
\medskip
\noindent{\bf c --- The level density parameter $a$ }\par
\smallskip
\indent For the nuclei considered here, Mathias \ref{14} has determined
the best values of $a$ that reproduce our combinatorial data. These
values are given in table 3, and agree to better than 10 \% with various
experimental determinations. \par
\medskip
\noindent{\bf d --- Accuracy of the level density }\par
\smallskip
\indent Let us analyse the origins of possible inaccuracies in the
determination of the level density. Considering only the exponential
term, we can write :
$$ { \Delta \rho \over \rho } = \sqrt{aU} \left( { \Delta a \over a } +
{ \Delta \delta \over U } \right) \eqno(7) $$
\indent The back-shift \delta gives an arror that is large at low
energies, but that vanishes for higher energies. For instance, with the
maximum error quoted previously (\ie $ 500 \ \keV $), the error arising
from the determination of \delta does not exceed 40 \% for an excitation
energy of $ 10 \ \MeV $ if $a \simeq 6 \ \MeV\sp{-1} $, which is a
typical value for the nuclei considered here. \par
\indent The relative error arising from the determination of $a$ remains
constant. Table 3 shows that $ \displaystyle{ \Delta a \over a } \simeq
0.15 $ leads to a factor of $2$ uncertainty in the level density at
$ 10 \ \MeV $. \par
\indent Hence, as long as $a$ remains the most difficult quantity to
predict for exotic nuclei, the level density will suffer an uncertainty
factor of, at least, two. \par
\vfill\eject
\medskip
\noindent{\bf e --- The spin distribution and the spin cut-off
parameter \sigma}\par
\smallskip
\indent The level density, as a function of energy and spin can be
written as :
$$ \rho(E,J) = f(U,J) \ \rho(E) \eqno(8) $$
\noindent where $ \rho(E) $ is the level density given by eq.$(2)$, and
$ f(U,J) $ is the spin distribution at energy $E$. It is expressed as :
$$ f(U,J) = { 2 J + 1 \over 2 \sigma^2 } \exp \left( - { J ( J + 1 )
\over 2 \sigma^2 } \right) \eqno(9) $$
\noindent where \sigma is the ``spin cut-off''. \par
\indent We have checked this well known gaussian distribution. As
expected, it fits very well the calculated spin distribution at any
energy, provided, of course, that a sufficient number of states exists
at the considered energy (see figure 2). \par
\indent The calculation of the corresponding value of \sigma can be
based on the fact that eq.$(9)$ presents a maximum at $ J \simeq
\sigma - { 1 \over 2 } $. Another method consists in the computation of
the average of $ J ( J + 1 ) $ over the spin distribution. In the limit
of a infinitely large number of spins, it leads to :
$$ \sum\sb{J=0}\sp{J=J\sb{max}} J(J+1) \ f(U,J) \simeq
\int\sb{0}\sp{\infty} J(J+1) f(U,J) dJ = 2 \sigma\sp{2} \eqno(10) $$
\indent Figure 3 presents values of \sigma for a large range of energy
bins. For our $ 10 \ \keV $ bin width value, we notice a large
dispersion of points which prevents any determination of a constant
value of \sigma (as guessed by ref. \ref{17}), or any check of the
``classical solid rotator law'' :
$$ \sigma^2 = { \theta \sb{rigid} \over \hbar\sp{2} } \sqrt{ U \over a }
\eqno{(11)} $$
$ \theta \sb{rigid} $ being the moment of inertia. \par
\indent Let us note that an increase of the width of the energy bins
reduces the dispersion of points, the new values of \sigma being
slightly greater than the previous ones. In fact, each of these new
values $ \sigma\sb{new} $ verifies the relation
$ \sigma\sb{new}\sp{2} = \sum\nolimits\sb{i} \sigma\sb{i}\sp{2} $,
where the summation runs over the old energy bins. \par
\indent Our determinations of \sigma can be compared with the constant
value :
$$ \sigma = ( 0.98 \pm 0.23 ) \  A \sp{ ( 0.29 \pm 0.06 ) } \eqno(12) $$
\noindent proposed in ref. \ref{17} from the data of the first excited
levels. For instance, we obtain $ \sigma \simeq 5. $ at $ 13.5 \ \MeV $
for for \nuc.Mn.50.  whereas eq.$(12)$ leads to
$ 1.84 \leq \sigma \leq 4.76 $ . \par
\indent Taking into account the fact that --1-- \sigma shows an overall
increase with energy, as perceptible on figure 3, --2-- eq.$(12)$ is
derived from the first excited levels and, --3-- our chosen energy of $
13.5 \ \MeV $ is far from the energies of the first levels (but it
ensures a very good gaussian distribution of spins) and we conclude
that, a constant value of \sigma cannot be rejected at least for
energies below about $ 20 \ \MeV $. \par
\indent Let us notice that, whatever the width of the bins can be (say
less than $ 500 \ \keV $, typically $ 100 \ \keV $, in order to have
enough energy points), there are some bins with \sigma values less than
the neighboring ones. Hence, the series of $ \{ \sigma \sb{i} \} $ is
not monotonic, and {\it all the narrower} the bins are. However, as
already stated, the overall trend is the increase of the upper and lower
limits of the \sigma values with energy (see figure 3). We do not
attempt here to evaluate quantitatively the variations of these upper
and lower bounds with energy or bin width (these two functions being
increasing functions in both variables). Instead, we limit ourselves to
the check of the $ U \sp{1/4} $ dependence predicted in eq.$(11)$. \par
\indent To do that, we have analysed the values \sigma for the energy
bins for which a {\bf new} maximum value in the spin distribution
(``yrast spin'') appears for the {\bf first} time. The interest of
such a procedure is that the resulting values of \sigma and energies
(represented by big black triangles on figure 3) are independent of the
bin width. \par
\indent Apart from one point, \sigma appears to increase with energy,
and a power law fit gives an exponent lying between $0.12$ and $0.17$ .
A Monte-Carlo calculation with bins of $ 1 \ \MeV $ up to an excitation
energy of $ 100 \ \MeV $ (see ref. \ref{19}) is in good agreement, with the
$ U \sp{1/4} $ law, as expected in the high energy limit. We conclude that
the discrepancy found here originates from the considered lower energies
which are necessary for astrophysical purposes. \par
\null
\bigskip
\beginsection{ 4 : THE PARITY DISTRIBUTION }\par
\medskip
\noindent{\bf a --- The problem of the equipartition of the parities}
\smallskip
\indent Following Ericson \ref{2}, it is classicaly assumed that the
equipartition of the parities holds at ``sufficiently'' high energy so
that we can write :
$$ \rho(E,J,\pi) = {1 \over 2 } \ f(U,J) \ \rho(U) \eqno(13) $$
\indent Further work has been devoted to that question (see, for
instance, ref. \ref{20}), but has been poorly investigated in the
framework of combinatorial level density calculations. \par
\indent The question of the parity distribution is important in a
variety of problems, and especially in the evaluation of the
transmission coefficients in the \gamma channel that appears in the
Hauser-Feschbach theory. \par
\medskip
\noindent{\bf b --- Results}\par
\smallskip
\indent The parity distribution at a given energy can be described in
terms of the asymmetry ratio :
$$ A(E) = { N\sp{+}(E) - N\sp{-}(E) \over N\sp{+}(E) + N\sp{-}(E) }
\eqno(14) $$
\noindent where $ N\sp{\pi} $ are the number of levels with parity
$ \pi = \pm 1 $ . Of course $ A(E=0) = \pi\sb{g} $ where $ \pi\sb{g} $
is the parity of the ground state, and the equipartition of parities is
realized when $ A(E) \simeq 0 $ . \par
\indent Figures 4a and 4b show the evolution of $A(E)$ with energy for
the even-even nucleus \nuc.Fe.56. and for the odd-odd nucleus
\nuc.Mn.50. . Obviously, $A(E)$ tends to $0$ more rapidly in the
latter case. Clearly, these figures show that the hypothesis of parity
equipartition is invalid below energies less than about $ 15 \ \MeV $,
in contrast with previous predictions. \par
\indent We now device a simple model in order to evaluate quantitatively
the critical energy above which $ A(E) \simeq 0 $. \par
\medskip
\noindent{\bf c --- A simple model}\par
\smallskip
\indent In the framework of the shell model and considering only
particle or particle-hole intrinsic excitations with no interaction
mixing, we introduce different energies for protons $(p)$ and neutrons
$(n)$ :  \par
\smallskip
\noindent \tnumt{1} $ D\sb{(n)} $ and $ D\sb{(p)} $ are the energy
differences of the last filled single particle level {\it splev}, called
for simplicity the Fermi {\it splev}, and the first {\it splev} with a
parity opposite to the Fermi {\it splev} , noted $ \pi \sp{0}
\sb{(n)} $ and $ \pi\sp{0}\sb{(p)} $ . $ D\sb{(n)} $ and $ D\sb{(p)} $
are counted positive for a single particle excitation, or negative for a
particle-hole excitation. \par
\indent We consider that, from $ 0 \;  \MeV $ to $ E \sb{1} = \min ( |
D\sb{(n)} | , | D\sb{(p)} | ) $, there is no way to form levels with a
parity different from the ground state one, $ \pi\sb{g} $ .
Hence, $ A(E) = \pi\sb{g} $ over this interval. This assumption is
strongly supported by the fact that, at low energy, the parity of
individual states is, most of the time, experimentally known and remains
constant for the first few states. Taking into account the parity of
$N$ and $Z$, we can improve somewhat the previous determination of $ E
\sb{1} $ by adding the corresponding pairing energies $ \Ep (n) $ or
$ \Ep (p) $ . \par
\smallskip
\noindent \tnumt{2} $ D\sb{(2n)} $ and $ D\sb{(2p)} $ are the energy
differences between Fermi {\it splev} and the first {\it splev} with a
parity equal to $ \pi \sp{0} \sb{(n)} $ and $ \pi\sp{0}\sb{(p)} $ that
is not in the same major shell as the Fermi {\it splev}. \par
\indent For excitation energies greater than $ E \sb{2} = \max ( |
D\sb{(2n)} | , | D\sb{(2p)} | ) $ , there are, at least, for protons and
for neutrons, two sets of {\it splev} with opposite parities which are
of comparable sizes. Therefore, we consider that $ A(E) \simeq 0 $ for
$ E \ge E \sb{2} $. Moreover, our simple model assumes that $A(E)$ goes
to $0$ linearly for $ E \sb{1} < E < E \sb{2} $ (see figure 5). \par
\indent Table 4 presents the values of $ E \sb{1} $ and $ E \sb{2} $
determined from our simple model. This model seems able to reproduce
the overall behavior of the parity distribution for most of the analyzed
nuclei. A few exceptions exist however :  \nuc.Ca.40., \nuc.Ti.44.,
\nuc.Ca.47. and \nuc.Ca.50. exhibit an additional ripple before $A(E)$
vanishes, but the physical interpretation of $ E \sb{1} $ and $ E \sb{2} $
remains the same. \par
\indent From a practical point of view, since $A(E)$ tends to $0$ when
the energy increases, as expected, we can consider that the
equipartition of parities is ``sufficiently'' achieved when the
$ | A(E) | < 0.25 $. With this convention, the equipartition of parity
is realized at an energy close to $ 15 \ \MeV $ for \nuc.Fe.56., and
between $ 12 \ \MeV $ and $ 14 \ \MeV $ \break for \nuc.Mn.50.~. Therefore,
the equipartition of parities is not realized at energies of astrophysical
interest at least, for light and moderately heavy nuclei. \par
\null
\bigskip
\beginsection{ 5 : CONCLUSIONS }\par
\smallskip
\indent We have presented an {\it ab initio} tool to compute the nuclear
level density in the framework of the combinatorial approach.  We are
able to predict the values of the level density parameter $a$ and we
give a prescription to evaluate the back-shift energy \hbox{\delta~.} \par
\indent We underline the difficulty encountered in the evaluation of the
spin cut-off parameter \sigma with energy. In fact, the choice of a
constant value has not to be rejected and the usual $ U \sp{1/4} $
law is not clearly established, at least at low energies.
We suggest to determine upper and lower bounds of \sigma versus energy,
according to the width of the bins. \par
\indent One of the major questions raised in this work concerns the
approach of the equipartition of the parity distribution with increasing
energy. We have introduced two critical energies, physically based on
the nuclear shell model, and we have presented a very simple model to
sketch the evolution of the asymmetry ratio of the parity distribution
with energy. \par
\bigskip
\noindent {\it Acknowledgements :  } It is a pleasure to thank Marcel
Arnould of the IAAG of ULB (Brussels) for initiating this work and the
PAI (Belgium) for a grant during the academic year 1987-88.  Georges
Audi (CSNSM, Orsay) is thanked to provide me with atomic masses and
other related quantities.  Thanks also to Claude Jacquemin and Jean
Treiner (IPN, Orsay) for helpful discussions.  I am also very much
indebted to Yvon G.  Biraud (DESPA, Meudon) for his help.  \par
\vfill \eject
\def\coupure{\hfill\break\indent}
\def\seealso{\hfill\break\indent{see also\ }}
\bigskip
\centerline{\bf References}
\null
\smallskip
\ref{ 1} A.H.Bethe \quad Phys. Rev. 50 (1936) 332 -- 341 \par
\smallskip
\ref{ 2} T.E.O.Ericson \quad Adv. Phys. 9 (1960) 425 -- 511 \par
\smallskip
\ref{ 3} J.R.Huizenga and L.G.Moretto \quad Ann. Rev. Nucl. Sci. 22 (1972) 427
-- 464 \par
\smallskip
\ref{ 4} M.Herman and G.Reffo \quad Comp. Phys. Comm. 47 (1987) 103 -- 111 \par
\smallskip
\ref{ 5} G.Maino and A.Ventura \quad Comp. Phys. Comm. 43 (1987) 303 -- 312
\par
\smallskip
\ref{ 6} M.Hillman and J.R.Grover \quad Phys. Rev. 185 (1969) 1303 -- 1319 \par
\smallskip
\ref{ 7} F.C.Williams \quad Nucl. Phys. A 133 (1969) 33 -- 49 \par
\smallskip
\ref{ 8} G.P.Ford \quad Nucl. Sci. Eng. 66 (1978) 334 -- 348 \par
\smallskip
\ref{ 9} F.Tondeur \quad Nucl. Phys. A 303 (1978) 185 -- 198 , \coupure
         Nucl. Phys. A 311 (1978) 51 -- 60 and
         Nucl. Phys. A 315 (1979) 353
\seealso Private communication \par
\smallskip
\ref{10} A.S.Jensen, P.G.Hansen and B.Jonson \quad Nucl. Phys. A 431 (1984) 393
-- 418 \par
\smallskip
\ref{11} P.M\"oller and J.R.Nix \quad Atom. Data Nucl. Data Tables 39 (1988)
213 -- 223 \par
\smallskip
\ref{12} P.M\"oller, W.D.Myers, W.J.Swiatecki and J.Treiner \quad \coupure
Atom. Data Nucl. Data Tables 39 (1988) 225 -- 233 \par
\smallskip
\ref{13} T.Tachibana, M.Uno, M.Yamada and S.Yamada \quad \coupure Atom. Data
Nucl. Data Tables 39 (1988) 251 -- 258 \par
\smallskip
\ref{14} Ph.Mathias \quad Rapport de stage de DEA (1989) unpublished \par
\smallskip
\ref{15} N.Cerf \quad Phys. Lett. B 268 (1991) 317 -- 322 \par
\smallskip
\ref{16} M.Pieper \quad Z. Phys. A 284 (1978) 203 -- 207 \par
\smallskip
\ref{17} T.VonEgidy, H.H.Schmidt and A.N.Behkami \quad Nucl. Phys. A 481 (1988)
189 -- 206 \coupure
         T.VonEgidy, A.N.Behkami and H.H.Schmidt \ Nucl. Phys. A 454 (1986) 109
-- 127 \par
\smallskip
\ref{18} A.H.Wapstra and G.Audi \quad Nucl. Phys. A 432 (1985) 55 -- 139
\seealso G.Audi Private communication (1991) \par
\smallskip
\ref{19} N.Cerf \quad private communication \par
\smallskip
\ref{20} M.Herman and G.Reffo \quad Phys. Rev. C 36 (1987) 1546 -- 1564 \par
\smallskip
\ref{21} S.E.Woosley, W.A.Fowler, J.A.Holmes and B.A.Zimmerman \quad \coupure
Atom. Data Nucl. Data Tables 22 (1978) 371 -- 441 \par
\smallskip
\ref{22} W.Dilg, W.Schantl, H.Vonach and M.Uhl \quad Nucl. Phys. A 217 (1973)
269 -- 298 \par
\smallskip
\vfill \eject
\null
\bigskip
\centerline{\bf Table 1}
\bigskip
$$ \vbox{
   \halign{
\hfil # \hfil \quad & \hfil # \hfil \quad & \hfil # \hfil \quad
& \hfil # \hfil \quad & \hfil # \hfil \quad & \hfil # \hfil \quad
& \hfil # \hfil \quad & \hfil # \hfil \quad & \hfil # \hfil \cr
\multispan6$\overbrace{\hbox{\hskip95truemm}}\sp{\displaystyle%%
( m_1 , m_2 , m_3 )}$ & $M$ & $E$ & level \cr
 $-5/2$ & $-3/2$ & $-1/2$ & $+1/2$ & $+3/2$ & $+5/2$    \cr
\noalign{\medskip}
\noalign{\noindent\hrule}
\noalign{\medskip}
  *   &   *   &   *   &       &       &       & $-9/2$ &   0    & (1)  \cr
  *   &   *   &       &   *   &       &       & $-7/2$ &   0    & (1)  \cr
  *   &   *   &       &       &   *   &       & $-5/2$ & $-E_p$ & (2') \cr
  *   &   *   &       &       &       &   *   & $-3/2$ & $-E_p$ & (2') \cr
  *   &       &   *   &   *   &       &       & $-5/2$ & $-E_p$ & (3') \cr
  *   &       &   *   &       &   *   &       & $-3/2$ &   0    & (1)  \cr
  *   &       &   *   &       &       &   *   & $-1/2$ & $-E_p$ & (2') \cr
  *   &       &       &   *   &   *   &       & $-1/2$ &   0    & (1)  \cr
  *   &       &       &   *   &       &   *   & $+1/2$ & $-E_p$ & (2') \cr
  *   &       &       &       &   *   &   *   & $+3/2$ & $-E_p$ & (2') \cr
      &   *   &   *   &   *   &       &       & $-3/2$ & $-E_p$ & (3') \cr
      &   *   &   *   &       &   *   &       & $-1/2$ & $-E_p$ & (3') \cr
      &   *   &   *   &       &       &   *   & $+1/2$ &   0    & (1)  \cr
      &   *   &       &   *   &   *   &       & $+1/2$ & $-E_p$ & (3') \cr
      &   *   &       &   *   &       &   *   & $+3/2$ &   0    & (1)  \cr
      &   *   &       &       &   *   &   *   & $+5/2$ & $-E_p$ & (2') \cr
      &       &   *   &   *   &   *   &       & $+3/2$ & $-E_p$ & (3') \cr
      &       &   *   &   *   &       &   *   & $+5/2$ & $-E_p$ & (3') \cr
      &       &   *   &       &   *   &   *   & $+7/2$ &   0    & (1)  \cr
      &       &       &   *   &   *   &   *   & $+9/2$ &   0    & (1)  \cr
 } } $$
\vfill
\noindent {\bf Table 1 :  } The $M-$configurations involved in the case
of a $ (\subsh.1.d.5.) \sp{3} $ configuration.  The column $M$ indicates
the resulting value of the spin-projection.  Here we find, as stated in
Mayer-Jensen'table, that such a configuration leads to one state for
each spin $ J = 9/2 $~, $ J = 5/2 $ and $ J = 3/2 $ .  \par
\indent The energy of each state, as evaluated by Ford's rule, is
displayed in the next column where $E_p$ is a given constant pairing
energy.  The last column indicates the identification of each state with
a corresponding spin.  Clearly, the states labelled $(1)$ gives,
unambiguously, a $ J = 9/2 $ spin with energy $ E = 0 $ (let us notice
that this level is ``incomplete'' in the sense that the states $ M = \pm
5/2 $ are not present).  The states labelled $(2')$ and $(3')$ give,
with some unspecified linear combinations, the level $(2)$ and $(3)$
with the same spin $ J = 5/2 $ with energy $ E = - E_p $.  \par
\vfill \eject
\null
\bigskip
\centerline{\bf Table 2}
\bigskip
$$\vbox{
\halign{
\hfil #     & \quad \hfil $#$
                  & \hfil $#$       \cr
Nucleus     & d_p   &  d_V  \cr
\noalign{\smallskip}
\noalign{\noindent ``Magic'' series \hfill\par}
\nuc.Ca.40. &  2.18 &  2.71 \cr
\nuc.Ti.44. &  2.11 &  3.45 \cr
\nuc.Ti.46. &  2.05 &  2.23 \cr
\nuc.Cr.48. &  2.05 &  2.58 \cr
\nuc.Cr.50. &  2.00 &  1.45 \cr
\noalign{\smallskip}
\noalign{\noindent Isobaric $ A = 47 $ series \hfill\par}
\nuc.K.47.  & -0.10 & -0.58 \cr
\nuc.Ca.47. &  0.07 & -0.40 \cr
\nuc.Sc.47. & -0.05 & -0.83 \cr
\nuc.Ti.47. &  0.03 & -0.44 \cr
\nuc.V.47.  & -0.03 & -1.36 \cr
\noalign{\smallskip}
\noalign{\noindent Isobaric $ A = 50 $ series \hfill\par}
\nuc.K.50.  & -1.20 & -1.87 \cr
\nuc.Ca.50. &  1.45 &  0.92 \cr
\nuc.Sc.50. & -1.66 & -1.87 \cr
\nuc.Ti.50. &  1.82 &  1.62 \cr
\nuc.V.50.  & -1.93 & -1.87 \cr
\nuc.Cr.50. &  2.00 &  1.45 \cr
\nuc.Mn.50. & -2.02 & -1.87 \cr
\nuc.Fe.50. &  2.00 &  1.52 \cr
\noalign{\smallskip}
\noalign{\noindent ``Ford'' series \hfill\par}
\nuc.Fe.56. &  1.88 &  1.12 \cr
\nuc.Co.59. & -0.04 & -0.58 \cr
\nuc.Ni.60. &  1.84 &  1.44 \cr
\nuc.Ni.62. &  1.76 &  1.72 \cr
\nuc.Cu.61. & -0.03 & -0.65 \cr
\nuc.Cu.63. & -0.04 & -0.53 \cr
}
}$$
\vfill
\noindent {\bf Table 2 :  } Values of the back-shift \delta as
determined in this work ($ col.  \ d_p $) and comparisons with other
available values :  from the phenomenological formulae of VonEgidy and
collaborators \ref{17} ( $ col.  \ d_V $ for their equation (11)~)~.
For \nuc.K.50.  the values of the (experimental) pairing energies are
not available and, therefore, the corresponding values for \delta by the
formulae of VonEgidy would not be calculable, but fortunately for an
odd-odd nucleus such as this one, the term coming from the neutron and
proton pairing vanishes.  Let us note, that as usually assumed, the
back-shift for odd $A$ nuclei is compatible with $0$ .  \par
\vfill \eject
{\vsize=28truecm
\null
%%%% \smallskip
\centerline{\bf Table 3}
\medskip
$$\vbox{
\halign{
\hfil #     & \quad \hfil #
                            & \hfil #
                                      & \hfil #
                                                & \hfil #
                                                          & \hfil #
                                                                    & # \hfil
 & # \hfil \cr
Nucleus     & $a_0$ & $a_2$ & $a_u$ & $a_d$ & $a_D$ & $a_V$ & $a_p$ \cr
\noalign{\smallskip}
\noalign{\noindent ``Magic'' series \par}
\nuc.Ca.40. &  6.87 &  6.15 &  3.7  &       &       &   3.6 &       \cr
\nuc.Ti.44. &  5.72 &  5.59 &       &       &       &       &       \cr
\nuc.Ti.46. &  5.35 &  5.31 &  5.0  &       &       &       &       \cr
\nuc.Cr.48. &  5.40 &  5.45 &       &       &       &       &       \cr
\nuc.Cr.50. &  5.22 &  5.09 &       &       &       &       &       \cr
\noalign{\smallskip}
\noalign{\noindent Isobaric $ A = 47 $ series \par}
\nuc.K.47.  &  5.42 &  5.28 &       &       &       &       &       \cr
\nuc.Ca.47. &  5.68 &  5.13 &       &       &       &       &       \cr
\nuc.Sc.47. &  4.93 &  4.96 &       &       &       &       &       \cr
\nuc.Ti.47. &  4.86 &  4.92 &  5.4  &  4.68 &  5.54 &       &  6.6  \cr
\nuc.V.47.  &  5.63 &  5.96 &  5.0  &       &       &       &       \cr
\noalign{\smallskip}
\noalign{\noindent Isobaric $ A = 50 $ series \par}
\nuc.K.50.  &  5.85 &  5.62 &       &       &       &       &       \cr
\nuc.Ca.50. &  5.97 &  5.49 &       &       &       &       &       \cr
\nuc.Sc.50. &  5.54 &  5.43 &       &       &       &       &       \cr
\nuc.Ti.50. &  5.46 &  5.07 &  5.5  &  5.53 &  5.26 &       &       \cr
\nuc.V.50.  &  5.17 &  5.25 &       &       &       &       &  6.3  \cr
\nuc.Cr.50. &  5.22 &  5.09 &       &       &       &       &       \cr
\nuc.Mn.50. &  5.35 &  5.65 &       &       &       &       &       \cr
\nuc.Fe.50. &  5.26 &  4.91 &       &       &       &       &       \cr
\noalign{\smallskip}
\noalign{\noindent ``Ford'' series \par}
\nuc.Fe.56. &  6.83 &       &  5.8  &       &       &       &       \cr
\nuc.Co.59. &  6.30 &       &  6.0  &  5.50 &  6.31 &       &       \cr
\nuc.Ni.60. &  6.77 &       &  6.6  &       &       &       &       \cr
\nuc.Ni.62. &  6.65 &       &  6.7  &  6.48 &  7.27 &       &       \cr
\nuc.Cu.61. &  6.63 &       &       &  5.03 &  5.99 &       &       \cr
\nuc.Cu.63. &  6.50 &       &  7.0  &  5.74 &  6.63 &       &       \cr
}
}$$
\vfill
\noindent {\bf Table 3 :  } Values of the level density parameter $a$
(in $ \MeV \sp{-1} $) as determined in this work ($ col.  \ a_0 $
without pairing, $ col.  \ a_2 $ with pairing, \cf eqs. $(1c)$ and
$(4)$ ) and comparisons with other available experimental values :  from
the publication \ref{21} \break ( $ col.  \ a_u $ , from the unpublished
Ph.D.  thesis of J.A.Holmes), from the work of Dilg and collaborators
\ref{22} , the two values ( $ col.  \ a_d $ ) and ( $ col.  \ a_D $ )
presented have been computed according to the assumption made about the
momentum of inertia (see the original reference for further details),
from the work of VonEgidy and collaborators \ref{16} ( $ col.  \ a_V $ )
, from the work of Pieper \ref{18} ( $ col.  \ a_p $ ) .  \par
}%%
\vfill \eject
\null
\bigskip
\centerline{\bf Table 4}
\bigskip
$$\vbox{
\halign{
\hfil #  & \quad \hfil $#$ &\ \hfil $#$ &\ \hfil $#$ &\ \hfil $#$ &\
\hfil $#$ &\ \hfil $#$ \cr
Nucleus &
         D\sb{(n)} &
         D\sb{(p)} &
                    D\sb{(2n)} &
                    D\sb{(2p)} &
                                E \sb{1} &
                                E \sb{2} \cr
\noalign{\smallskip}
\noalign{\noindent ``Magic'' series \hfill\par}
\nuc.Ca.40. &  3.82 &  3.66 &  14.25 & 12.68 &  3.66 & 14.25 \cr
\nuc.Ti.44. & -4.01 & -3.82 & -15.73 & 10.88 &  3.82 & 15.73 \cr
\nuc.Ti.46. & -4.09 & -3.87 & -15.83 & 12.43 &  3.87 & 15.83 \cr
\nuc.Cr.48. & -4.17 & -3.96 & -15.83 & 11.68 &  3.96 & 15.83 \cr
\nuc.Cr.50. & -4.25 & -4.01 & -15.95 & 13.18 &  4.01 & 15.95 \cr
\noalign{\smallskip}
\noalign{\noindent Isobaric $ A = 47 $ series \hfill\par}
\nuc.K.47.  & -4.17 &  3.98 & -16.16 & 14.25 &  3.98 & 16.16 \cr
\nuc.Ca.47. & -4.14 &  3.87 & -16.06 & 14.00 &  3.87 & 16.06 \cr
\nuc.Sc.47. & -4.12 & -3.87 & -15.93 & 14.37 &  3.87 & 15.93 \cr
\nuc.Ti.47. & -4.12 & -3.90 & -15.91 & 13.27 &  3.90 & 15.91 \cr
\nuc.V.47.  & -4.13 & -3.91 & -15.84 & 12.02 &  3.91 & 15.84 \cr
\noalign{\smallskip}
\noalign{\noindent Isobaric $ A = 50 $ series \hfill\par}
\nuc.K.50.  &  6.51 &  3.90 &  15.51 &  14.04 & 3.90 & 15.51 \cr
\nuc.Ca.50. &  6.52 &  3.86 &  15.85 &  13.88 & 3.86 & 15.85 \cr
\nuc.Sc.50. &  6.66 & -3.94 &  16.28 & -15.06 & 3.94 & 16.28 \cr
\nuc.Ti.50. & -4.26 & -3.99 & -16.08 & -15.24 & 3.99 & 16.08 \cr
\nuc.V.50.  & -4.27 & -4.01 & -16.09 &  14.41 & 4.01 & 16.09 \cr
\nuc.Cr.50. & -4.25 & -4.01 & -15.95 &  13.18 & 4.01 & 15.95 \cr
\nuc.Mn.50. & -4.26 & -4.03 & -15.91 &  12.05 & 4.03 & 15.91 \cr
\nuc.Fe.50. & -4.26 & -4.05 & -15.81 &  10.94 & 4.05 & 15.81 \cr
\noalign{\smallskip}
\noalign{\noindent ``Ford'' series \hfill\par}
\nuc.Fe.56. &  5.85 & -4.09 & 15.77 &  14.98 &  4.09 & 15.77 \cr
\nuc.Co.59. &  5.20 & -4.04 & 15.22 & -15.01 &  4.04 & 15.22 \cr
\nuc.Ni.60. &  5.10 & -4.08 & 15.15 & -15.06 &  4.08 & 15.15 \cr
\nuc.Ni.62. &  2.97 & -4.10 & 12.87 & -14.88 &  2.97 & 14.88 \cr
\nuc.Cu.61. &  5.13 &  5.80 & 15.16 &  11.47 &  5.13 & 15.16 \cr
\nuc.Cu.63. &  2.90 &  5.61 & 12.78 &  12.49 &  2.90 & 12.78 \cr
}
}$$
\vfill
\noindent {\bf Table 4 : } Values of the parameters $ D\sb{(n)} $ ,
$ D\sb{(p)} $ , $ D\sb{(2n)} $ , $ D\sb{(2p)} $ , $ E \sb{1} $ and
$ E \sb{2} $ used in our simple model (see text \S 4-c). \par
\vfill \eject
\null
\bigskip
\centerline{\bf FIGURE CAPTIONS}
\bigskip
\def\Fig#1{{\noindent \bf Figure #1 : }}
\Fig{1a} Rough data of density level (with summation over the spins and
parities) for the nucleus of \nuc.Fe.56. (with no-pairing energy). \par
\smallskip
\Fig{1b} The same data of figure 1a but after treatment by a running
mean method over 51 points (\ie $ 500 \;  \keV $).  The fitted curve
with $ a = 6.83 \;  \MeV \sp{-1} $ is shown.  Let us note that the level
density is shown and not the cumulative number of levels.  \par
\medskip
\Fig{2} A typical spin distribution (solid curve) as calculated for the
\nuc.Mn.50.  nucleus at an energy of $ 13.5 \ \MeV $ for parity plus
levels (we recall that the bin has a width of $ 10 \ \keV $) compared to
the theoretical one (dashed curve) calculated with $ \sigma = 5.  $ .
Here, the total number of levels is less than 1000 (about 860) and we
can conclude than the gaussian law is valid for a total number of levels
of this order of magnitude.  \par
\medskip
\Fig{3} Values of the spin cut-off parameter \sigma for every bins of
energy, from $ 0 \;  \MeV $ to $ 28 \;  \MeV $ by step of $ 10 \;  \keV
$ excluding bins with less than 100 levels, with parity $ \pi = + 1 $
for the \nuc.Mn.50. nucleus.  For the big solid triangles, see text (\S
3-e).  \par
\medskip
\Fig{4a} Evolution of the parity distribution function $ A(E) $ (see
text for its definition) with the energy for the \nuc.Fe.56.  nucleus.
\par
\smallskip
\Fig{4b} The same as figure (4a) for the \nuc.Mn.50. nucleus. \par
\medskip
\Fig{5} Our simple modelisation of the parity distribution function
$A(E) $ with the two key-energies $ E \sb{1} $ and $ E \sb{2} $.  \par
\medskip
\vfill \eject
\bye